\documentclass[amsfonts,onecolumn,nofootinbib]{revtex4-1}
\usepackage{graphicx}
\usepackage{amsmath}
\usepackage{amsfonts}
\usepackage[dvips]{epsfig}
\usepackage[cm]{fullpage}
\usepackage{bm}
\usepackage[active]{srcltx}
\usepackage{epsf}

\begin{document}

\author{Shmuel Marcovitch and Benni Reznik}
\title{Correlation preserving map between bipartite states and temporal evolutions}
\affiliation{School of Physics and Astronomy,
Raymond and Beverly Sackler Faculty of Exact Sciences,
Tel-Aviv University, Tel-Aviv 69978, Israel.}
\begin{abstract}
The Hilbert space formalism of quantum theory manifests a map between bipartite states and time evolutions,
known as Jamio{\l}kowski isomorphism.
We extend this map in a physical setting to prove the equality of spatial correlations in bipartite systems
and temporal correlations in local systems.
We show that these correlations can be observed using weak measurements.
This result has several practical and conceptual implications such as manifestation of state independent decoherence,
the correspondence between Bell
and Leggett-Garg inequalities, multipartite systems, the statistical properties of evolutions in large systems,
and computational gain, in evaluation of spatial correlations in large systems.
\end{abstract}
\maketitle


\section{Introduction}
The quantum mechanical nature of bipartite states is well known.
Among bipartite states the maximally entangled ones manifest maximal nonlocality, a unique property of quantum theory.
These states may be regarded as possessing the highest degree of quantumness.
One may expect that in the time domain unitary evolutions would manifest the highest degree of quantumness too, as they describe
evolutions without environment interference.
But how do we test the quantum mechanical nature of evolutions?
Using tomography of states at different times the evolution and environment traces can be tracked given a known pure state to begin with.
However, this method is indirect.
In particular, with tomography the evolution of the maximally mixed state cannot be tested.

The temporal correlations of observables before and after the tested dynamics, however,
provide a direct test of the evolution.
In this letter we construct a one-to-one Jamio{\l}kowski map \cite{jam} between the space of bipartite systems
$\rho_{AB}\in H_A\otimes H_B$
and the space of time evolutions transforming systems from Hilbert spaces $H_A$ to $H_B$.
In this map spatial correlations of two separated operators in bipartite systems
precisely coincide with the temporal correlations of the mapped operators in local systems.
Thus, the entanglement between $A$ and $B$ is mapped to a correlation between the past and the future,
which characterize the evolutions of systems and their quantum mechanical nature.

The suggested map provides interesting physical consequences, as shown
schematically in table 1. In particular, the maximally entangled states are mapped to unitary evolutions --
making explicit the expectation that unitary evolutions in dynamical processes are as maximal entangled states
in bipartite systems.
Non-maximally entangled states correspond to evolutions under the influence of selective measurements wherein the environment
is observed and one particular outcome is selected.
Specifically, pure product states correspond to selective projector measurements.
Mixed bipartite systems are mapped to mixtures of the corresponding evolutions.
Closed systems with non-selective environment correspond to bipartite states in which the reduced density matrices
are the maximally mixed ones.

\begin{table}[ht]
\caption{One to one map between time evolutions and bipartite states} 
\centering 
\begin{tabular}{c c } 
\hline\hline 
Temporal & Spatial \\ [0.5ex] 
(assuming the system is maximally mixed $\rho_S=I/d_A$)& \\
\hline
Pure evolution &  Pure bipartite system  \\
$M=\sqrt{d_A}\sum_{ij}\alpha^{*}_{ji}|i\rangle\langle j|$ where $\rho_S\rightarrow M\rho_S M^\dagger$
\footnote{Notice we employ a normalization slightly different from that in the usual Kraus representation (see section II).}& $|\psi_{AB}\rangle=\sum_{ij}\alpha_{ij}|i\rangle\otimes|j\rangle$ \\
$\text{Tr}[M^\dagger M\rho_S]=1$ & $\langle\psi_{AB}|\psi_{AB}\rangle=1$ \\
Unitary evolution $M=U,\  UU^\dagger=I$ & Maximally entangled $|\psi_{AB}\rangle$ \\ 
Selective projector $M=\sqrt{d_A}|i\rangle\langle j|$ &  Product state $|\psi_{AB}\rangle=|i\rangle\otimes|j\rangle$\\
\hline
Mixed evolution $(\sum p_\mu=1)$ & Mixed bipartite system $(\sum p_\mu=1)$\\
$\{M_\mu,\ p_\mu\}:\ \rho_S\rightarrow \sum_\mu p_\mu M_\mu\rho_S M_\mu^\dagger$ & $\{|\psi^\mu_{AB}\rangle,\ p_\mu\}:\ \rho_{AB}=\sum_\mu p_\mu |\psi^\mu_{AB}\rangle\langle\psi^\mu_{AB}|$ \\
Non-selective environment $\sum_\mu p_\mu M_\mu^\dagger M_\mu=I$ & $\rho_A=I/d_A$, $\rho_B=I/d_B$\\
[1ex] 
\hline\hline
\end{tabular}
\label{table1} 
\end{table}


It may be argued that in general two sequentially measured operators do not commute and
effect each other due to the uncertainty principle.
Therefore, the temporal correlations are limited to the formalistic analysis,
and cannot be measured.
However, as is well known, there is a trade-off between
the accuracy of the measurement and the disturbance
caused to the system \cite{von}.
The limit in which individual measurements provide vanishing information gain
was first analyzed by Aharonov {\it et. al.} \cite{weak} 
and was termed {\it weak measurements}.
Since weak measurements only slightly disturb the systems,
they provide a non-destructive and operational method for measuring temporal correlations by which
the effect of evolutions can be measured directly.

One may have noticed that the system in the temporal setting has no counterpart in the spatial setting --
it is the evolution that maps to a bipartite system.
In table 1 we indeed assume a trivial initial system in the temporal setting $\rho^{in}_S=I/d_A$.
To overcome this we extend the mapping to include any initial system in the temporal setting.
Surprisingly, this state is mapped to a local final state in the spatial setting,
that is a post-selection of only one of the parties.
Pleasantly, a final state in the temporal setting then corresponds to a final state
of the second party in the spatial setting.
This construction is illustrated in fig. 1 (a) and (b).



The paper is organized as follows. In the following section we set the ground for the mapping and
provide preliminary definitions regarding generalized time evolutions.
In section III we define the map between time evolutions and bipartite states.
In this section we provide the main result on the equality of temporal and spatial correlations.
In addition we show in section III that the temporal (and spatial) correlations can be measured by utilizing weak measurements.
In section IV we generalize the map to include initial and final states.
In section V we prove our results.
In section VI we suggest several implications of the suggested map: manifestation of state independent decoherence,
the correspondence between Bell \cite{bell}
and Leggett-Garg \cite{leggett} inequalities, partitions of multipartite states, the statistical properties of evolutions in large systems,
and computational gain, in evaluation of spatial correlations in large systems.
We conclude in section VII.

\begin{figure}[ht]
\center{
\includegraphics[width=3.8in]{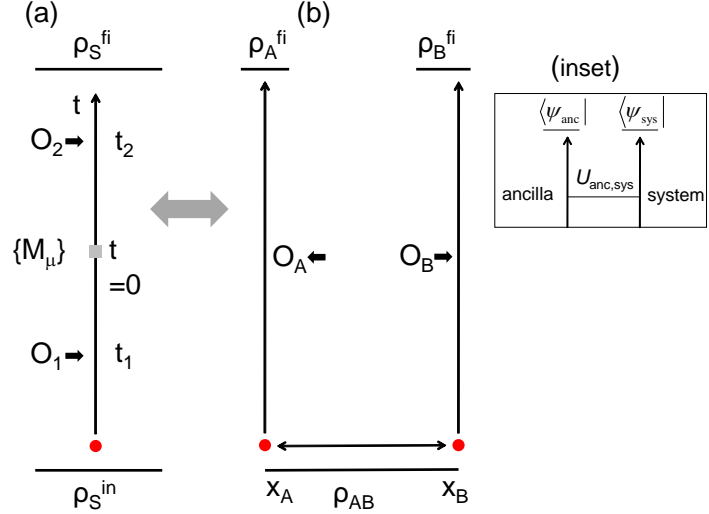}
\caption
{Mapping of bipartite states to time evolutions.
(a) A local state with initial and final states $\rho_S^{\text{in}}$ and $\rho_S^{\text{fi}}$ respectively is weakly measured at $t_1<0$ and $t_2>0$
by $O_1$ and $O_2$ respectively,
where the system undergoes an instantaneous evolution given by operators $\{M_\mu\}$ at $t=0$.
(b) $A$ and $B$ share a bipartite state $\rho_{AB}$ and weakly measure it by $O_A$ and $O_B$ respectively.
The system has local final states $\rho_A^{\text{fi}}$ and $\rho_B^{\text{fi}}$.
$\rho_{AB}$, $O_A(x_A)$, $O_B(x_B)$, $\rho_A^{\text{fi}}$ and $\rho_B^{\text{fi}}$ are mapped to
$\{M_\mu\}$, $O_1(t_1)$, $O_2(t_2)$, $\rho_S^{\text{in}}$ and $\rho_S^{\text{fi}}$, respectively.
\\Inset. Realization of post-selection to a mixed state by interaction with ancilla and post selecting both the system and the ancilla
to pure states.
}
\label{fig1}
}
\end{figure}

\section{Preliminary definitions}
To set the ground for the mapping let us first discuss generalized time evolutions.
The evolution of a system $\rho_S$, subject to interaction with a larger system,
is most generally described as a completely positive map given by
Kraus operators $\{\mathcal{K}_\mu\}$:
\begin{equation}
\label{K0}
\rho_S\rightarrow\sum_{\mu=1}^N \mathcal{K}_\mu\rho_S \mathcal{K}_\mu^{\dagger},
\end{equation}
where
\begin{equation}
\label{constraint}
\sum_\mu \mathcal{K}_\mu^{\dagger} \mathcal{K}_\mu=I.
\end{equation}
Beyond the trivial unitary evolutions $\rho_S\rightarrow U\rho_S U^\dagger$, Kraus operators describe evolutions
due to the interaction with an environment.

It will be convenient to use a set of Kraus operators $M_\mu$ defined as
\begin{equation}
M_\mu\equiv\frac{\sqrt{d_A}\mathcal{K}_\mu}{\sqrt{\text{Tr}(\mathcal{K}_\mu^\dagger\mathcal{K}_\mu})}
\equiv\frac{\mathcal{K}_\mu}{p_\mu},
\end{equation}
which satisfy the normalization
\begin{equation}
\label{norm0}
\text{Tr}(M_\mu^\dagger M_\mu)=d_A.
\end{equation}
Then Eqs. (\ref{K0},\ref{constraint}) become
\begin{eqnarray}
\label{changeM1}&&\rho_S\rightarrow \sum_{\mu=1}^N p_\mu M_\mu\rho_S M_\mu^{\dagger},\\
\label{changeM2}&&\sum_\mu p_\mu M_\mu^{\dagger} M_\mu=I.
\end{eqnarray}
Eqs. (\ref{changeM1},\ref{changeM2}) (or \ref{K0},\ref{constraint}) provide the most general description of the evolution of a system which is part of a larger {\it closed} system.

The above formalism, however, does not describe the scenario in which an
external observer measures the environment and selects a certain outcome.
In case a single outcome corresponding to $M_\mu$ is selected, the evolution
is given by $\rho_S\rightarrow M'_\mu \rho_S {M'}_\mu^\dagger$
where $M'_\mu$ is normalized
to preserve the trace of the density matrix
$M'_\mu=M_\mu/\sqrt{\text{Tr}( M_{\mu}^{\dagger} M_{\mu} \rho_S)}$.
In the even more general case of selective measurements
more than a single selection can be made.
In this case we have $\sum_\mu p_\mu=1$ where
$p_\mu$ is the probability to select $M_\mu$, and the evolution of the system is given by
$\rho_S\rightarrow\sum_\mu p_\mu M'_\mu\rho_S {M'}_\mu^\dagger$.
The normalization of the $M_\mu$ then changes according to
\begin{equation}
\label{normalization}
M'_\mu=\frac{M_\mu}{\sqrt{\text{Tr}[\sum_{\nu} p_{\nu} M_{\nu}^{\dagger} M_{\nu} \rho_S]}},
\end{equation}
in order to preserve the trace of $\rho_S$.
Note that in the non-selective setting,
the constraint imposed by Eq. (\ref{changeM2}) is satisfied
and the probabilities $p_\mu$ are fixed,
whereas the selective setting corresponds to the case Eq. (\ref{changeM2}) is not satisfied
and the probabilities $p_\mu$ are arbitrary.
A set of normalized operators and probabilities $\{M_\mu,p_\mu\}$ is sufficient to describe the evolution,
where Eq. (\ref{changeM1}) should be taken with the $M'_\mu$ according to the normalization in Eq. (\ref{normalization}).
Notice that in case $\rho_S=I/d_A$, $M_\mu=M'_\mu$, which is the case illustrated in table 1.



Clearly, a set of operators and probabilities $\{M'_\mu,\ p_\mu\}$ is not unique.
It can be transformed to a set $\{N'_\nu, \ q_\nu\}$
with a unitary transformation $U_{K\times K}$, producing the un-normalized operator
\begin{equation}
\label{nu}
\tilde{N}'_\nu=U_{\nu\mu}\sqrt{p_\mu}M'_\mu.
\end{equation}
The corresponding probability is then $q_\nu=\text{Tr}({\tilde{N}}^{'\dagger}_\nu \tilde{N}'_\nu\rho_S)$
and the normalized operator $N'_\nu=\tilde{N}'_\nu/\sqrt{q_\nu}$.
Again one can represent the set $\{N'_\nu, q_\nu\}$ by the canonical set
$\{N_\nu, q_\nu\}$ in which the $N_\nu$ are normalized according to Eq. (\ref{norm0}).


\section{The map}
In the temporal setting we assume an initially prepared system $\rho_S^{\text{in}}$ with dimension $d_A$ and internal Hamiltonian $H_0$
is subject to an evolution described by operators $M_\mu$
(as normalized in Eq. \ref{normalization}),
which without loss of generality we take as instantaneous at time $t=0$.
In addition, for clarity we assume the internal Hamiltonian in the spatial setting vanishes.

Following a formal definition of the map between time evolutions and bipartite states.
Any pure bipartite state is mapped to a single
(normalized) operator by
\begin{equation}
\label{M}
|\psi_\mu\rangle=\sum_{ij}\alpha^\mu_{ij}|i\rangle\otimes|j\rangle\quad\Leftrightarrow\quad M_\mu=\sqrt{d_A}\sum_{ij}\alpha^{\mu*}_{ji}|i\rangle\langle j|.
\end{equation}
where $d_A$ is the dimension of $H_A$, $1\leq i\leq d_A$, and $d_B$ of $H_B$, $1\leq j\leq d_B$.
The map extends to mixed states/evolutions by convex combinations:
\begin{equation}
\label{M2}
\rho_{AB}=\sum_\mu p_\mu |\psi_\mu\rangle\langle\psi_\mu|\quad\Leftrightarrow\quad \rho_S\rightarrow \sum_\mu  p_\mu M'_\mu \rho_S M_\mu^{'\dagger}.
\end{equation}
Note that the map does not define a spatial correspondence to the initial state
in the temporal setting $\rho_S^{\text{in}}$.
We shall assume (for now) that this state is maximally mixed $\rho_S=I/d_A$.
Nontrivial initial (and final) states are discussed in the following.

In tables 1 the correspondence between evolutions and states is given for several cases.
In particular, it is evident from Eq. (\ref{M}) that maximally entangled states are mapped to unitary evolutions.
\\\\{\bf Lemma 1}. {\it Non-selective environment in the temporal setting maps to a state $\rho_{AB}$ in which each of the reduced states is maximally
mixed $\rho_A=I/d_A$, $\rho_B=I/d_B$.}
Lemma 1 strengthens the physical essence of the map. The usual non-selective evolution is one in which either
we are ignorant regarding the effect of the environment, time has directionality and the future cannot be known in advance,
or in case the evolution is purely unitary.
This is mapped in the spatial setting to the scenario in which locally each party is maximally ignorant regarding her state.
In particular, the reduced state of the pure maximally entangled state
is maximally mixed.
\\\\{\bf Temporal correlations.}
Before stating our main result we would like to discuss temporal correlations
\footnote{Throughout the paper the term {\it correlation} corresponds to the expectation of the product of operators,
sometimes regarded as correlator, without subtracting the first moments as required in the statistical definition of the term.}.
Let there be two Hermitian operators
$O_1(t_1)$, $O_2(t_2)$
given in the Heisenberg representation
\begin{equation}
\label{heisenberg}
O_i(t_i)=e^{iH_0 t_i} O_ie^{-iH_0 t_i}\quad (\text{taking }\hbar=1).
\end{equation}
Clearly, the effect of operators $\mathcal{K}_\mu$ on the evolution of observable $O$ is exactly as that on the density operator:
$O\rightarrow\sum_\mu \mathcal{K}_\mu O \mathcal{K}_\mu^\dagger$.
Therefore, the most straightforward definition of the temporal correlation of two operators $O_1$ and $O_2$ at instances
$t_1$ and $t_2$ respectively is given by
\begin{equation}
E[O_2(t_2)O_1(t_1)]=\frac{1}{d_A}\text{Tr}\left[O_2(t_2)\,\sum_\mu p_\mu M'_\mu \,O_1(t_1) \,M_\mu^{'\dagger}\right],
\end{equation}
where $d_A^{-1}$ is a normalization factor.
\\{\bf Theorem 1}. {\it Let there be two Hermitian operators in the
temporal setting $O_1(t_1)$, $O_2(t_2)$, $t_1<0<t_2$, and two operators in the spatial setting $O_A(x_A)$, $O_B(x_B)$, such that
$O_1(t_1)=O_A(x_A)$, $O_2(t_2)=O_B^{\text{T}}(x_B)$.
Given the mapping defined in Eqs. (\ref{M}, \ref{M2}),
the temporal and spatial correlations equal:}
\begin{equation}
\label{formal}
\frac{1}{d_A}\text{Tr}\left[O_2\,\sum_\mu p_\mu M'_\mu \,O_1 \,M_\mu^{'\dagger}\right]=\text{Tr}\left[\,O_A\,\otimes \,O_B \,\rho_{AB}\,\right],
\end{equation}
where we omit the $t$ and $x$ parameters from now on.
This mapping is illustrated in figure 1, where for now we assume that in the temporal setting the state is $\rho_S=I/d_A$
and no final states are assumed.
\\ {\bf Corollary 1}. {\it The expectation values of the single operators equal as well:}
\begin{equation}
E(O_1)=E(O_A),\quad E(O_2)=E(O_B).
\end{equation}

Eq. (\ref{formal})
is symmetric to the exchange of indices $A$ and $B$, given
that we take $M_\mu^{'\text{T}}$ instead of $M'_\mu$ and reverse the time direction $t\rightarrow -t$.
That is, given a point in spacetime the mapping is symmetric under simultaneous time and space reflections.

Note that representation transformation of the evolution with unitary $U$,
as given in Eq. (\ref{nu})
maps in the spatial setting to the same representation transformation.
That is, the un-normalized $\tilde{N}'_\nu$ maps to the un-normalized state
$|\tilde{\phi}^\nu_{AB}\rangle=U_{\nu\mu}\sqrt{p_\mu}|\psi^\mu_{AB}\rangle$.
The normalized
$N'_\nu=\tilde{N}'_\nu/\sqrt{\text{Tr}(\tilde{N}_\nu^{'\dagger} \tilde{N}'_\nu \rho_S)}$
with probability
$q_\nu=\text{Tr}(\tilde{N}_\nu^{'\dagger} \tilde{N}'_\nu \rho_S)$
maps to
$|\phi_\nu\rangle=|\tilde{\phi}_\nu\rangle/\sqrt{\langle\tilde{\phi}_\nu|\tilde{\phi}_\nu\rangle}$
with probability
$q_\nu=\langle\tilde{\phi}_\nu|\tilde{\phi}_\nu\rangle$,
and the evolution
$\rho_S\rightarrow \sum_\nu q_\nu N'_\nu\rho_S N_\nu^{'\dagger}$ maps to
$\rho_{AB}=\sum_\nu q_\nu|\phi^\nu_{AB}\rangle\langle\phi^\nu_{AB}|$.

\subsection{Measuring the temporal correlation with weak measurements}

Let us now utilize {\it weak measurements} \cite{weak} to show that the temporal correlation in the LHS of Eq. (\ref{formal}) can be measured.
We assume that the system is measured weakly (and instantaneously) at $t_1<0$ and $t_2>0$
by operators $O_1(t_1)$ and $O_2(t_2)$ with two pointer readings $q_1$ and $q_2$ respectively,
as illustrated in fig. 1(a).
\\{\bf Lemma 2.} {\it
The correlation of the instruments' pointers is given by:}
\begin{equation}
\label{temporal}
E(q_1^{\text{weak}} q_2^{\text{weak}} )=\frac{1}{2}\text{Tr}\left[O_2\,\sum_\mu p_\mu M'_\mu\{O_1,\rho_S^{\text{in}}\}M_\mu^{'\dagger}\right],
\end{equation}
which includes both selective and non-selective measurements.
Eq. (\ref{temporal}) reduces to the temporal correlation in the LHS of Eq. (\ref{formal})
given that $\rho_S^{\text{in}}=I/d_A$.
See related results in the context of unitary evolutions 
for correlations of two-level system with continuous weak measurements \cite{korotkov}, in the context of post-selection \cite{steinberg,jozsa,stein2} and of two sequential measurements \cite{johansen,ours}.

In the spatial setting we assume
that the initially prepared bipartite system $\rho_{AB}$ is measured by
parties $A$ and $B$ with operators $O_A$ and $O_B$ with pointers $q_A$ and $q_B$ respectively, as illustrated in fig. 1(b).
Note that the spatial correlation in the RHS of Eq. (\ref{formal}) can be measured with regular strong measurements.
But an immediate consequence of Eq. (\ref{temporal}) is that given weak measurements too
\begin{equation}
\label{spatial}
E(q_A^{\text{weak}} q_B^{\text{weak}}) = \text{Tr}\left[O_A \otimes O_B\, \rho_{AB}\right].
\end{equation}
Weak measurements in the spatial setting will become essential in the following generalization of the map to include post-selection of final states.

In passing we note that a sequence of three weak measurements 
$O_1$, $O_2$ and $O_3$ at times $t_1<t_2<t_3$ where $M=I$, yield
\begin{equation}
\label{tripartite}
E(q_1^{\text{weak}}q_2^{\text{weak}}q_3^{\text{weak}})=\frac{1}{4}\text{Tr}\left[O_3,\{O_2,\{O_1,\rho_S^{\text{in}}\}\}\right],
\end{equation}
In contrast to weak measurements at two times, Eq. (\ref{tripartite}) implies that the correlation of three depends on their order,
in contradiction with the multipartite spatial scenario.

\section{Generalizing the map to include initial and final states}
Remarkably, the mapping above can be generalized to the case that the initial state in the temporal setting is not maximally mixed.
In this case the state corresponds to a local final state of one of the parties in the spatial setting.
In addition, a final state in the temporal setting corresponds to a local final state of the second party in the spatial setting.
Explicitly, in the case of an initial state $\rho_S^{\text{in}}$ with dimension $d_A$ and a final state $\rho_S^{\text{fi}}$ with dimension $d_B$ in the temporal setting,
Lemma 2 (Eq. \ref{temporal}) generalizes to
\\{\bf Lemma 3.}
\begin{equation}
\label{post}
E(q_1^{\text{weak}} q_2^{\text{weak}}) = \frac{1}{4}\text{Tr}\left[\rho_S^{\text{fi}}\left\{O_2,\sum_\mu p_\mu M'_\mu\{O_1,\rho_S^{\text{in}}\}M^{'\dagger}_\mu\right\}\right],
\end{equation}
where
\begin{equation}
\label{C2}
M'_\mu=\frac{M_\mu}{\sqrt{\text{Tr}\left[\rho_S^{\text{fi}}\sum_{\nu} p_{\nu} M_{\nu} \rho_S^{\text{in}} M^{\dagger}_{\nu}\right]}}.
\end{equation}
Physically, one can prepare a final mixed state by post-selecting the state and an ancilla
as described in the inset of fig. 1 and in the proof of the lemma in section V.
Note that having no final state is equivalent to having a maximally distributed final state $\rho_S^{\text{fi}}=I/d_B$.

In the spatial setting, having final states $\rho_A^{\text{fi}}$ in $A$ and $\rho_B^{\text{fi}}$ in $B$
generalizes Eq. (\ref{spatial}) using Eq. (\ref{post}) to
\begin{equation}
\label{spatial2}
E(q_A^{\text{weak}} q_B^{\text{weak}}) = \frac{\text{Tr}\left[\rho_A^{\text{fi}} \otimes \rho_B^{\text{fi}}\left\{I_A\otimes O_B,\{O_A\otimes I_B,\rho_{AB}\}\right\}\right]}
{4\text{Tr}\left[ (\rho_A^{\text{fi}}\otimes \rho_B^{\text{fi}} )\rho_{AB}\right]}.
\end{equation}
Note that given final states in the spatial setting weak and strong measurements provide different results,
where the mapping applies only to the weak measurement regime.
\\{\bf Theorem 2.} {\it Given that $\rho_S^{\text{in}} = \rho_A^{\text{fi}}$,
and $\rho_S^{\text{fi}}=\rho_B^{fi\,\text{T}}$,
Theorem 1 generalizes to $E(q_1^{\text{weak}}q_2^{\text{weak}})=E(q_A^{\text{weak}} q_B^{\text{weak}})$ or
explicitly},
\begin{equation}
\frac{1}{4}\text{Tr}\left[\rho_S^{\text{fi}}\left\{O_2,\sum_\mu M'_\mu\{O_1,\rho_S^{\text{in}}\}M^{'\dagger}_\mu\right\}\right]=
\frac{1}{4}\frac{\text{Tr}\left[\rho_A^{\text{fi}} \otimes \rho_B^{\text{fi}}\left\{I_A\otimes O_B,\{O_A\otimes I_B,\rho_{AB}\}\right\}\right]}
{\text{Tr}\left[ (\rho_A^{\text{fi}}\otimes \rho_B^{\text{fi}} )\rho_{AB}\right]}.
\end{equation}
The map is fully illustrated in figure 1.

\section{Proofs of the results}
We proceed by proving the results to the cases
including initial and final states: Theorem 2, Lemma 2 and Lemma 3,
where reduction to the case of no(/totally mixed) initial and final
states in the temporal setting and no final states in the spatial
setting (Theorem 1) is straightforward.
Lemma 1 on the correspondence between the usual non-selective environment in the temporal setting
and having maximally mixed reduced density matrices in the spatial setting, can be proved straightforwardly, by
imposing $\sum p_\mu M^\dagger_\mu M_\mu = I$ and computing the corresponding $\rho_A=\text{Tr}_B\rho_{AB}$.
$O_i$ in the temporal setting is given by Eq. (\ref{heisenberg}).
\\{\bf Proof of Theorem 2. } 
Let us first show the correspondence for a pure bipartite state $|\psi\rangle$, which is mapped to a single operator $M$ (with $p=1$).
We show the equality of the temporal and spatial denominators $D_T$, $D_S$ and nominators $N_T$ and $N_S$ of Eqs. (\ref{post}) and (\ref{spatial2}) respectively.
From Eqs. (\ref{C2},\ref{M}) up to a factor of $4$:
\begin{equation}
\begin{split}
&D_T=\text{Tr}\left[\rho_S^{\text{fi}} M \rho_S^{\text{in}} M^{\dagger}\right]
=\sum_{i,j,k,l} \alpha_{ij}\alpha^*_{kl}\rho_{S_{ki}}^{\text{in}}\rho_{S_{ij}}^{\text{fi}},
\\&D_S=\text{Tr}\left[(\rho_A^{\text{fi}}\otimes \rho_B^{\text{fi}})|\psi\rangle\langle \psi |\right]
=\sum_{i,j,k,l} \alpha_{ij}\alpha^*_{kl}\rho_{A_{ki}}^{\text{fi}}\rho_{B_{lj}}^{\text{fi}},
\\&N_T\!=\!\text{Tr}\!\left[\!\rho_S^{\text{fi}}\!\{O_2,M\{O_1,\rho_S^{\text{in}}\}\! M^\dagger \}\!\right]\!
=\!\sum_{i,j,k,l}\!\alpha_{ij}\alpha^*_{kl}\{\rho_S^{\text{in}},O_1\}_{ki}\{\rho_S^{\text{fi}},O_2\}_{jl},
\\&N_S\!=\!\text{Tr}\!\left[\!(\rho_A^{\text{fi}}\!\otimes \!\rho_B^{\text{fi}})\!\{I_A\otimes O_B,\{O_A\otimes I_B,|\psi\rangle\langle\psi|\}\!\}\!\right]\!
=\!\sum_{i,j,k,l}\alpha_{ij}\alpha^*_{kl}\{\rho_A^{\text{fi}},O_A\}_{ki}\!\{\rho_B^{\text{fi}},O_B\}_{jl},
\end{split}
\end{equation}
where we use the notation $A_{ki}=\langle k|A|i\rangle$ for matrix elements.
In correspondence with the mapping defined in Theorems (1,2), $D_T=D_S$ and $N_T=N_S$.
By proving $D_S=D_T$ we have explicitly confirmed that the mapping corresponds to Jamio{\l}kowski isomorphism \cite{jam}.
To extend to a set of operators $\{M'_\mu,\ p_\mu\}$, note that $D_T$, $D_S$, $N_T$, $N_S$ become now a convex combinations of $p_\mu$,
which respects their equality. This concludes the proof of Theorems 2. $\square$
\\\\{\bf Proof of Lemma 2. }
Observables $O_1,\ O_2$ are measured sequentially on system $\rho_S^{\text{in}}$
at times $t_1, t_2$ where $t_1<0<t_2$.
In addition, $\rho_S^{\text{in}}$ evolves at $t=0$
with operators $\{p_\mu,M'_\mu\}$.
The von-Neumann interaction measurement corresponding to $O_1$ and $O_2$ is
\begin{equation}
H_{int}=\delta(t-t_1)p_1 O_1+\delta(t-t_2)p_2 O_2,
\end{equation}
where $[q_i,p_i]=i (\hbar=1)$.
We assume identical initial Gaussian wavepackets $\phi(q_1)$ and $\phi(q_2)$ for the pointers: 
\begin{equation}
\rho_i=\phi(q_i)\phi(q'_i)=\int dq_i dq'_i \sqrt{\frac{\epsilon}{2\pi}}e^{-\epsilon (q_i^2+q_i^{'2})/4}, \quad(i=1,2).
\end{equation}

The initial state of the system and the apparatuses $\rho_S^{\text{in}}\otimes\rho_1\otimes\rho_2$,
evolves to
\begin{equation}
U_2\left[\sum_\mu p_\mu M'_\mu(U_1 \,\rho_S^{\text{in}} \otimes \rho_1\otimes\rho_2\, U_1^{\dagger})M_\mu^{'\dagger}\right]U_2^{\dagger},
\end{equation}
where
$U_i=e^{-i p_i O_i}$. 
Each operation of $p$ yields an order of $\epsilon$, where in the limit of weak measurements $\epsilon\to0$.
By expanding $U_i$ to second order ($i=1,2$)
$U_i=1-i p_iO_i-\frac{1}{2}p_i^2O_i^2+o(\epsilon^3)$,
one can compute the composite state of the system and pointers:
\begin{equation}
\label{expand}
\begin{split}
& \rho_S^{\text{in}}\otimes\rho_1\otimes\rho_2\ \rightarrow \ \sum_\mu p_\mu M'_\mu\rho_S^{\text{in}} \rho_1\rho_2M_\mu^{'\dagger}
 - \sum_\mu p_\mu M'_\mu\left\{O_1,\rho_S^{\text{in}}\right\} M_\mu^{'\dagger}\phi'(q_1)\phi(q_1')\rho_2
 +\frac{1}{2}\sum_\mu p_\mu M'_\mu\left\{O_1^2,\rho_S^{\text{in}}\right\} M_\mu^{'\dagger}\phi''(q_1)\phi(q_1')\rho_2
\\& \!-\! \left\{\!O_2,\!\sum_\mu p_\mu M'_\mu \rho_S^{\text{in}} M_\mu^{'\dagger}\!\right\}\!\phi'(q_2\!)\phi(q_2')\rho_1
\!+\!\frac{1}{2}\!\left\{\!O_2^2,\!\sum_\mu p_\mu M'_\mu \rho_S^{\text{in}}M_\mu^{'\dagger}\!\right\}\!\phi''\!(q_2\!)\phi(q_2')\rho_1
\\&\!+\!\left\{\!O_2,\!\sum_\mu p_\mu M'_\mu\!\left\{\!O_1,\rho_S^{\text{in}}\!\right\}\! M_\mu^{'\dagger}\!\right\}\!\phi'(q_1\!)\phi(q_1')\phi'(q_2\!)\phi(q_2').
\end{split}
\end{equation}
First note that since
$\int q\phi^2(q)dq=0$ and $\int\phi(q)\phi'(q)dq=0$,
all terms in Eq. (\ref{expand}) except the last do not contribute.
By tracing out the system $\rho_S^{\text{in}}$ and using $\int q \phi(q)\phi'(q)dq=-1/2$,
we obtain
\begin{equation}
E(q_1^{\text{weak}} q_2^{\text{weak}}) = \frac{1}{4}\text{Tr}\left[\left\{O_2,\sum_\mu p_\mu M'_\mu\left\{O_1,\rho_S^{\text{in}}\right\} M_\mu^{'\dagger}\right\}\right],
\end{equation}
which coincide with Eq. (\ref{temporal}).
\\{\bf Proof of Lemma 3 (Eq. \ref{post}). }
Preparation of a mixed state is realized by projecting a system to a pure state $|\psi_S^{\text{in}}\rangle$
which then interacts with an ancilla in a known state. 
Correspondingly, post selection to a mixed state $\rho_S^{\text{fi}}$ is realized in the reversed order:
\begin{equation}
\rho_S^{\text{fi}}=U_{int}^{\dagger}|0_{anc}\rangle\langle0_{anc}|\otimes|\psi_S^{\text{fi}}\rangle\langle\psi_S^{\text{fi}}|U_{int},
\end{equation}
(as illustrated in the inset of figure 1).
Then the proof follows the same steps as that of Lemma 1 where instead of tracing out the system,
one projects the system to the final state and renormalizes the remaining state.
In case $M=I$ the normalization yields a factor of
\begin{equation}
\frac{1}{\text{Tr}[\langle 0_{anc}|\langle \psi_S^{\text{fi}}|U_{int}\rho_S^{\text{in}}\otimes I_{anc} U_{int}^{\dagger}|\psi_S^{\text{fi}}\rangle|0_{anc}\rangle]}=\frac{1}{\text{Tr}[\rho_S^{\text{in}}\rho_S^{\text{fi}}]}.
\end{equation}
The generalization to arbitrary evolution is straightforward.
Note that Wizeman \cite{wiseman} analyzed a similar case
for a single weak measurement.

\section{Implications}
The suggested mapping has several interesting implications which we discuss in some detail.
\subsection{State independent decoherence}
A common framework of decoherence deals with the transition of a state to a one with higher level of mixing.
By the suggested mapping
one can distinguish decoherence of states from {\it decohering dynamics}.
Decohering dynamics can be observed by detecting the temporal decay of correlations in case of non exact unitary evolution,
even on the {\it maximally distributed mixed state}. This might be important in problems of
tomography of dynamics in which the initial state cannot be prepared.

Let us discuss the simplest manifestation, in which
we can write the evolution of a system $\rho_S(t)$ according to the Lindblad master equation \cite{lindblad}:
\begin{equation}
\label{lindblad}
\dot{\rho}_S=-i [\hat{H},\rho_S(0)]+\sum_k \left[\hat{L}_k \rho_S(0) \hat{L}_k^\dagger-\frac{1}{2}\{\rho_S(0),\hat{L}_k^\dagger\hat{L}_k\}\right],
\end{equation}
where $\hat{L}_k$ are Lindblad operators satisfying $\hat{K}=-\frac{1}{2}\sum_k\hat{L}_k^\dagger \hat{L}_k$
such that $\hat{K}$ is Hermitian.
For simplicity, consider a single Linbdlad operator $L=\sigma_z$ which starts operating at $t=0$ where the free Hamiltonian vanishes.
This simplifies Eq. (\ref{lindblad}) at $t>0$ to
\begin{equation}
\label{l2}
\dot{\rho}_S=\sigma_z \rho_S \sigma_z-\rho_S.
\end{equation}
In addition we assume $\rho_S=I/2$, thus it does not change in time.

We manifest the decoherence model by studying a Leggett-Garg inequality \cite{leggett}, the Bell inequality \cite{bell} in time.
Specifically, we are looking on the temporal analog of CHSH inequality \cite{chsh} (Eq. 2b in \cite{leggett}), where both 
the spatial and temporal inequalities are bounded by $2\sqrt{2}$ \cite{tsirelson} in quantum theory and by 2 in classical theories.
In a variant of this test to weak measurements \cite{mizel} a single observer measures four observables $O_1,\ O_2, O_3,\ O_4$ at instances $t_1<t_2<t_3<t_4$ and look on
a combination of their correlations:
\begin{equation}
B_{LG}=\frac{1}{2}\text{Re}[O_1 O_3+O_1 O_4 + O_2 O_3 - O_2 O_4].
\end{equation}
We choose to measure
$\sigma_z(t=t_1)$, $\sigma_x(t=t_2)$, $\sigma_{\pi/4}(t=t_3)$ and $\sigma_{3\pi/4}(t=t_4)$,
where $$\sigma_{\frac{\pi}{4},\frac{3\pi}{4}}=\frac{1}{\sqrt{2}}\left(\sigma_x\pm\sigma_z\right).$$
Given the trivial evolution $M=I$, $B_{LG}=2\sqrt{2}$ for any initial state.
But now let us apply the evolution imposed in Eq. (\ref{l2}) from $t=0$, where
we assume $t_1,t_2<0$ and $t_3,t_4>0$ so the effect of decoherence 
takes place in the last two measurements.
Transforming to the Heizenberg representation each Pauli operator $\sigma$ transforms as
$\dot{\sigma}=\sigma_z\sigma\sigma_z-\sigma$.
Therefore off-diagonal terms decays in time as $\sigma_{12}(t)=\sigma_{12}e^{-\sqrt{2}\sigma_{12}t}$.
Using Eq. (\ref{temporal}) we obtain
\begin{equation}
B_{LG}=\frac{\sqrt{2}}{2}\Big(2+e^{-\sqrt{2}t_3}+e^{-\sqrt{2}t_4}\Big).
\end{equation}
Given $t_3=t_4$, $B_{LG}$ decays to 2 as $t_3\sim 0.623$.
By this simple application decoherence is manifested on the observables due to the non-unitary evolution,
whereas the state $\rho_S$ is constantly the maximally mixed state.
\subsection{The correspondence between Bell and Leggett-Garg inequalities} 
The temporal inequalities suggested by Leggett and Garg \cite{leggett} have the same bounds as the corresponding spatial Bell inequalities \cite{bell}.
For example, CHSH inequality \cite{chsh} and the corresponding temporal inequality (Eq. 2b in \cite{leggett}) are bounded by $2\sqrt{2}$ \cite{tsirelson}.
In a previous paper \cite{ours} we have shown that Bell's inequalities can be maximally violated using weak measurements even if
all observables are measured for each member of the ensemble.
A similar result for Leggett-Garg inequalities was given in \cite{mizel,jordan1}.
By our mapping the correspondence between these two types of inequalities becomes clear.
In particular, Leggett-Garg inequalities are distinguished from the Bell inequalities as their maximal violation depends only on
the measured observables and not on the state of the system.
By the mapping above we see that this is a consequence of unitary evolutions which correspond to maximally entangled bipartite states.
By having non unitary evolutions Leggett-Garg inequalities are less violated.

In particular, the CHSH inequality is maximally violated by the maximally entangled (Bell) state, for example
$|\psi\rangle=\frac{1}{\sqrt{2}}(|00\rangle+|11\rangle)$,
which is mapped in the temporal setting to $M=I$.
The case in which no post-selection is performed in the spatial case corresponds to having the
maximally mixed initial state $\rho_S=I/2$ in the temporal test. The $2\sqrt{2}$ bound on Leggett-Garg inequality is then obtained by half the trace of the matrix
\begin{equation}
B_{LG}=\text{Re}[\sigma_x\sigma_{\pi/4}+\sigma_x\sigma_{3\pi/4}+\sigma_z\sigma_{\pi/4}-\sigma_z\sigma_{3\pi/4}]=2\sqrt{2}I.
\end{equation}
Since post-selection of a single party does not change the bound on CHSH inequality,
the same bound is obtained for any initial state in the temporal setting.

Partial violation of CHSH inequality is obtained with non-maximally entangled pure or mixed states.
The unentangled pure product states correspond to selective projectors imposing "disentanglement" between the past
and the future.
The most familiar example regarding mixed states is Werner states \cite{werner}.
Werner states are convex combinations of the four Bell states, 
which correspond to a convex combination of the four unitary operations which constitute a basis.


Since our mapping is exact all the results concerning bipartite Bell inequalities are valid in the corresponding temporal inequalities.
An example is the anomaly of nonlocality in bipartite systems \cite{anomaly},
in which there are generalized Bell inequalities
that are not maximally violated by the maximally entangled state.
In particular, let us explore Collins-Gissin-Linden-Massar-Popescu (CGLMP) inequality \cite{g_bell2} (see also \cite{g_bell2add}).
This generalized Bell inequality corresponds to a setting in which every party measures two observables,
with $d$ outcomes (instead of the dichotomic observables in the CHSH test).
In a local hidden variable model CGLMP inequality is bounded by $2$ for any $d$.
It is maximally violated by a non-maximally entangled state for $d>2$.
One can explicitly show that the same anomaly appears in the temporal setting,
where maximal violation is obtained with the corresponding non-unitary evolution.

Let us show this result explicitly. 
As shown in \cite{anomaly} the Bell operator corresponding to CGLMP inequality in case $d=3$
is
\begin{equation}
B_{CGLMP}=\left(\begin{array}{ccccccccc}0 & 0 & 0 &0 & \frac{2}{\sqrt{3}} & 0 & 0 & 0 & 2\\
0 & 0 & 0 &0 & 0 & \frac{2}{\sqrt{3}} & 0 & 0 & 0\\
0 & 0 & 0 &0 & 0 & 0 & 0 & 0 & 0\\
0 & 0 & 0 &0 & 0 & 0 & 0 & \frac{2}{\sqrt{3}} & 0\\
\frac{2}{\sqrt{3}} & 0 & 0 &0 & 0 & 0 & 0 & 0 &\frac{2}{\sqrt{3}}\\
0 & \frac{2}{\sqrt{3}} & 0 & 0 & 0 & 0 & 0 & 0 & 0\\
0 & 0 & 0 &0 & 0 & 0 & 0 & 0 & 0\\
0 & 0 & 0 & \frac{2}{\sqrt{3}} & 0 & 0 & 0 & 0 &0\\
2 & 0 & 0 &0 & \frac{2}{\sqrt{3}} & 0 & 0 & 0 & 0\end{array}
\right).
\end{equation}
For the maximally entangled state $|\psi\rangle=\frac{1}{\sqrt{3}}(|00\rangle+|11\rangle+|22\rangle)$
the violation is $\langle\psi|B_{CGLMP}|\psi\rangle=\frac{4}{9}(3 + 2 \sqrt{3})\sim2.87293$.
However, the maximal eigenvalue of $B_{CGLMP}$ is $\frac{1}{3} (3 + \sqrt{33})\sim2.91485$
with eigenstate $|\psi_m\rangle=\frac{1}{\sqrt{n}}(|00\rangle+\gamma|11\rangle+|22\rangle)$,
where $\gamma=(\sqrt{11}-\sqrt{3})/2$ and $n=2+\gamma^2$.

Correspondingly, in the temporal setting we study the case of maximally mixed initial state
$I/3$, as no post-selection is assumed in the spatial setting.
In case the evolution
corresponds to the maximally entangled state $M=I$, we obtain
\begin{equation}
B^3_{LG}=\left(\begin{array}{ccc}
2+\frac{2}{\sqrt{3}} & 0 & 0\\0 & \frac{4}{\sqrt{3}} & 0\\0 &0 & 2+\frac{2}{\sqrt{3}}
\end{array}\right),
\end{equation}
where
$\text{Tr}[B_{LG}]/3=\frac{4}{9}(3 + 2 \sqrt{3})$.
Now let us set the non-unitary evolution corresponding to $|\psi_m\rangle$
\begin{equation}
M^m=\eta\left(\begin{array}{ccc}1 & 0 & 0\\ 0 & \gamma & 0 \\0 & 0 & 1\end{array}\right),
\end{equation}
where $\eta=\sqrt{3/(11/2 -\sqrt{33}/2)}$ such that $\text{Tr}[M^\dagger M]/3=1$.
Then
\begin{equation}
B^{3m}_{LG}=\left(\begin{array}{ccc}\frac{1}{2}+\frac{7}{2\sqrt{33}} & 0& 0\\0 & \frac{4}{\sqrt{33}}& 0\\0 & 0 &\frac{1}{2}+\frac{7}{2\sqrt{33}}\end{array}\right),
\end{equation}
such that $\text{Tr}[B^m_{LG}]/3=\frac{1}{3} (3 + \sqrt{33})\sim2.91485$ with the same bound as in the spatial setting.

Interestingly, given a maximally entangled state in the spatial setting,
post-selection of a single party produces higher violation of CGLMP inequality for $d>2$
(in contrast to the $d=2$ case).
This can be easily seen by checking the temporal setting and having a state $\rho_S$ different from the maximally mixed state $I/3$.
(Since the evolution is unitary $M=I$, one need not renormalize $B^3_LG$.)
Maximal violation corresponds to the maximal eigenvalue of $B^3_{LG}$: $2+\frac{2}{\sqrt{3}}\sim3.1457$,
where the corresponding eigenstates are $|0\rangle$ or $|2\rangle$.
This result shows that in the usual context of Leggett-Garg inequalities
in which only unitary evolutions are considered,
the quantum-mechanical bound is in general distinguished from the one in the corresponding Bell inequality.
The ordinary bound of Leggett-Garg inequalities with a unitary evolution and an arbitrary initial state
corresponds to the bound on the corresponding Bell inequality using the maximally entangled state given post-selection of a single party.

Another example is the $I_{3322}$ inequality suggested by Collins and Gisin \cite{collins}.
In \cite{collins} a (non-optimal) quantum mechanical bound of $\frac{1}{4}$ is found using maximally entangled Bell states.
However, in \cite{i3322_2,i3322_3} it is shown that the violation is higher for non-maximally entangled states,
thus providing another example of the anomaly of nonlocality.

Now let us discuss the corresponding temporal inequality.
In case the state is the maximally mixed and the evolution is unitary with $M=I$,
it can be shown that $I_{3322}\leq\frac{1}{4}$. The bound is satisfied by taking the operators $A_1,A_2,A_3,B_1,B_2,B_3$ along the $xy$ plane
with $\phi^A_1=0,\phi^A_2=\pi/3,\phi^A_3=-\pi/3,\phi^B_1=\pi/3,\phi^B_2=0,\phi^B_3=2\pi/3$.
In the spatial case the bound of $\frac{1}{4}$ does not depend on the dimension for maximally entangled states.
Therefore, the same characteristic is mapped to the temporal test.

\begin{table}[ht]
\caption{Anomaly of nonlocality in the temporal and spatial settings.} 
\centering 
\begin{tabular}{c c c} 
\hline\hline 
Temporal & & Spatial \\ [0.5ex] 
\hline
$\rho_S=I/d$. LG inequality maximally  & $\Leftarrow$ & Bell inequality maximally \\
violated with non-unitary evolution & & violated with non-maximally entangled state  \\
$\Downarrow$ & & $\Uparrow$ \\
$\rho_S$ general. LG inequality maximally & $\Rightarrow$ & Local post-selection. \\
violated with unitary evolution & & Bell inequality maximally violated \\
& & with maximally entangled state \\[1ex] 
\hline 
\end{tabular}
\label{table1} 
\end{table}

For arbitrary initial state the optimal violation is $B^{LG}_{3322}\leq\frac{3\sqrt{5}}{8}-\frac{1}{2}\sim0.3385$,
which is satisfied already at $d=2$ with
$\phi^A_1=0,\phi^A_2=2\pi/5,\phi^A_3=-2\pi/5,\phi^B_1=\pi/5,\phi^B_2=-\pi/5,\phi^B_3=3\pi/5$.
Through the map $I_{3322}$ has the same bound 0.3385 in case the state is a maximally entangled one and one of the parties
post-selects to the initial state of the temporal setting. This can be explicitly shown with the maximally entangled state $\frac{1}{\sqrt{2}}(|00\rangle+|11\rangle)$,
where $\phi^A_1=0,\phi^A_2=2\pi/5,\phi^A_3=-2\pi/5,\phi^B_1=-\pi/5,\phi^B_2=\pi/5,\phi^B_3=-3\pi/5$.

We believe our mapping provides a new perspective on the anomaly of nonlocality,
as illustrated in table 2. In the regular setting the anomaly of nonlocality
maps in the temporal setting to maximally violating the corresponding Leggett-Garg inequality
with a maximally mixed state and non-unitary evolution.
But then if we allow any initial state in the temporal setting, maximal violation
is obtained with a unitary evolution.
This maps back in the spatial setting to having a local post-selection and maximal
violation of the Bell inequality with the maximally entangled state.


\subsection{Multipartite systems}
The suggested map applies to any splitting of a multipartite state to two parts
corresponding to the past and future of the system in the temporal setting.
For example, let us have a tripartite system $|\psi\rangle=\sum_{ijk}\alpha_{ijk}|a_i\rangle\otimes
|b_j\rangle\otimes|c_k\rangle$,
where $|a_i\rangle$, $|b_j\rangle$, $|c_k\rangle$ correspond to bases of the Hilbert spaces of parties
$a,b,c$.
Consider a splitting of $|\psi\rangle$ to two parts, {\it e.g.} $a,b$ in one and $c$ in the other.
Through the mapping $|\psi\rangle$ translates to an evolution of a system lying in the Hilbert space of $a,b$
into the Hilbert space of $c$.
This evolution can be seen as a gate operating on $a,b$ (and an environment).
It translates the entanglement of $a,b$ with respect to the $c$ into a temporal correlation.

\subsection{Statistical characteristics of large systems}
In the work by Hayden {\it et. al.} \cite{popescu_thermo}
correlation properties of random high-dimensional
bipartite pure systems were examined in the context of the Haar measure.
They showed that there exist large subspaces in which almost all pure states
are close to maximally entangled.
Through the mapping the space of bipartite states maps to "pure evolutions" on local systems where a pure evolution
corresponds to a single $M$ operator. 
By mapping the Haar measure to the temporal setting, it follows that
there exist large subspaces of evolutions in which all pure evolutions are close to unitary ones.

\subsection{Computational gain}
In numerical computations of two point correlation functions of bipartite states, one can instead compute the temporal correlations.
For example, given $d_A=d_B=N$, instead of manipulating $N^2\times N^2$ matrices, one can use only $N\times N$ matrices.

\section{Discussion and conclusions}
Interestingly, genuine multipartite entanglement has no natural generalization through the map we suggest,
as can be seen from the structure of Eq. (\ref{M}).
The map is sensitive to the uniqueness of the one dimensional time with respect to space.
Time can only be bisected once to unordered parts,
whereas space may be sectioned into many parts with no internal order.
In particular, a tripartite state has no inherent order, due to the causal structure of space,
whereas three events in time have an internal order and are not symmetric under all permutations (see Eq. \ref{tripartite}).
Another feature of multipartite states which is not satisfied in the temporal setting is
the monogamy of entanglement \cite{wootters2}.
If two qubits $A$ and $B$ are maximally entangled, they cannot be correlated at all with a third qubit $C$.
In the temporal case, however, we can choose $M=I$.
Then any pair of instances among $t_1$, $t_2$, $t_3$ {\it etc.} is maximally correlated.

A notion of {\it entanglement in time} was introduced in a different context by Brukner {\it et. al.} \cite{vedral},
who analyze correlations of successive $\pm1$ strong measurements.
These temporal correlations violate Leggett-Garg inequalities \cite{leggett},
the Bell inequalities \cite{bell} in time.
Brukner {\it et. al.} show that there are no genuine multi-time correlations 
and that the monogamy of spatial correlations is violated in the temporal setting.
However, there are crucial differences between temporal correlations of strong and
weak measurements as
correlations of successive strong measurements do not depend on the state and are a particular feature of spin observables.
The suggested mapping does not apply for strong measurements.

We note that Leifer has shown an extension of Jamio{\l}kowski isomorphism
to include two POVMs before and after the evolution in the temporal setting and in parallel in the spatial setting \cite{leifer},
to manifests the correspondence of no cloning/no broadcasting theorems
and monogamy of entanglement
(see also \cite{griffiths}).
Jamio{\l}kowski mapping has also been used to analyze channel capacity \cite{frank,pablo,cirac,Nowakowski}.
Manifestation of "superposition of unitary operations" is given in \cite{benni}.
\\\\To conclude,
space and time are distinguished in the formalism of quantum theory.
A system that is separated in two parts of space is described by a positive semi-definite operator that lies in a tensor product of two Hilbert spaces.
The time evolution of a local system is described by a trace preserving complete positive map from one Hilbert space to another Hilbert space.
Mathematically, one can define a Jamio{\l}kowski
map 
between the space of
bipartite states and the space of time evolutions, which is
defined by the Hilbert-Schmidt scalar product.
We extended Jamio{\l}kowski map in a physical setting
in which the entanglement of bipartite states finds exact correspondence with temporal correlations between the past and future.
We use the tool of weak measurements to show that these correlations can be observed.
In particular, one can show that the map can be tested with a single ion and two ions
in an ion-trap respectively,
where instead of the displacement operator the pointer observables correspond to the phonon number operator.
By having an exact mapping between spatial and temporal correlations,
non-relativistic quantum-mechanics manifests a structural unification of time and space.

\acknowledgments
We are deeply grateful to Y. Aharonov whose insights initiated this work. We also thank A. Botero and P. Skrzypczyk. 
This work has been supported by the Israel Science Foundation grant number 920/09, the German-Israeli foundation, and the European Commission (PICC).

\end{document}